\documentclass{PoS}

\usepackage{epsf}
\usepackage{latexsym,times,mathptmx,amssymb,euscript}
\usepackage[numbers,sort&compress]{natbib}

\usepackage[italian,english]{babel}
\usepackage[T1]{fontenc}
\usepackage{graphicx,xcolor}
\usepackage[format=hang,font=small]{caption}
\usepackage{amsmath}
\usepackage{amsfonts}
\usepackage{amssymb}
\usepackage{booktabs}

\newcommand{\beq}[1]{
\begin{equation}\label{#1}}
\newcommand{\eeq}{\end{equation}}
\newcommand{\bea}[1]{
\begin{eqnarray}\label{#1}}
\newcommand{\eea}{\end{eqnarray}}
\newcommand{\barr}{
\begin{array}}
\newcommand{\earr}{\end{array}}

\title{Inclusive production of two rapidity-separated heavy quarks as a probe of BFKL dynamics}

\ShortTitle{Inclusive production of two rapidity-separated heavy quarks}

\author{Andr\`ee Dafne Bolognino\\
  Universit\`a della Calabria \& INFN - Gruppo collegato di Cosenza, \\
  I-87036 Arcavacata di Rende, Cosenza, Italy\\
  E-mail: \email{ad.bolognino@unical.it}}

\author{Francesco Giovanni Celiberto\\
  Universit\`a di Pavia \& INFN - Sezione di Pavia, I-27100 Pavia, Italy\\
  E-mail: \email{francescogiovanni.celiberto@unipv.it}}

\author{Michael Fucilla\\
  Universit\`a della Calabria, I-87036 Arcavacata di Rende, Cosenza, Italy\\
  E-mail: \email{mike.fucilla@libero.it}}

\author{Dmitry Yu. Ivanov\\
  Sobolev Institute of Mathematics, 630090 Novosibirsk, Russia\\
  Novosibirsk State University, 630090 Novosibirsk, Russia\\
  E-mail: \email{d-ivanov@math.nsc.ru}}

\author{Beatrice Murdaca\\
  Universit\`a della Calabria \& INFN - Gruppo collegato di Cosenza,\\ 
  I-87036 Arcavacata di Rende, Cosenza, Italy\\
  E-mail: \email{beatrice.murdaca@fis.unical.it}}

\author{\speaker{Alessandro Papa}\\
  Universit\`a della Calabria \& INFN - Gruppo collegato di Cosenza,\\
  I-87036 Arcavacata di Rende, Cosenza, Italy\\
  E-mail: \email{alessandro.papa@fis.unical.it}}

\abstract{The inclusive photoproduction of two heavy quarks, separated by a
  large rapidity interval, is proposed as a new channel for the manifestation
  of the Balitsky-Fadin-Kuraev-Lipatov (BFKL) dynamics. The extension to the
  hadroproduction case is also discussed.}

\FullConference{XXVII International Workshop on Deep-Inelastic Scattering and Related Subjects - DIS2019\\
		8-12 April, 2019\\
		Torino, Italy}

\begin{document}

\section{Introduction}

{\em Semihard processes} ($s \gg Q^2 \gg \Lambda_{\rm QCD}^2$, with
$s$ the squared center-of-mass energy, $Q$ the process hard scale
and $\Lambda_{\rm QCD}$ the QCD mass scale) represent a challenge for
high-energy QCD. Fixed-order perturbative calculations miss the effect of
large energy logarithms, which must be resummed to all orders. The
theoretical tool for this resummation is the BFKL approach~\cite{BFKL},
valid both in the leading logarithmic approximation (LLA) (all terms
$(\alpha_s\ln(s))^n$) and in the next-to-LLA (NLA) (all terms 
$\alpha_s(\alpha_s\ln(s))^n$). In this approach, the (possibly differential)
cross section factorizes into two process-dependent impact factors and a
process-independent Green's function.
Only a few impact factors have been calculated with next-to-leading order 
accuracy: parton to parton~\cite{partonIF}, parton to forward jet~\cite{jet},
parton to forward hadron~\cite{Ivanov:2012iv}, $\gamma^*$ to light vector
meson~\cite{IKP04} and $\gamma^*$ to $\gamma^*$~\cite{gammaIF}.
They were used to build predictions for a few {\em exclusive} processes:
$\gamma^* \gamma^*$ to two light vector mesons~\cite{mesons} and the
$\gamma^* \gamma^*$ to all~\cite{photons}, which can be studied in
future high-energy linear colliders. They enter, however, a lot
of {\em inclusive} processes, accessible at LHC: Mueller-Navelet jet
production~\cite{MN}, three and four jets, separated in rapidity~\cite{3-4jet},
two identified rapidity-separated hadrons~\cite{hh}, forward identified light
hadron and backward jet~\cite{Bolognino:2018oth}, forward $J/\Psi$-meson and
backward jet~\cite{Boussarie:2017oae}, forward Drell-Yan pair and backward
jet~\cite{Golec-Biernat:2018kem}.
Here we present another possible BFKL probe: the inclusive
production of two heavy quarks, separated in rapidity, in $\gamma \gamma$
collisions ({\em photoproduction}),
\beq{process}
\gamma(p_1) + \gamma(p_2)  \longrightarrow Q(q_1) 
\ + X \ + Q(q_2) \;,
\eeq
where $Q$ here stands for a $c$- or $b$-quark
(see Fig.~\ref{fig:process}(left)). This process can be studied either
at $e^+ e^-$ or in nucleus-nucleus colliders via the interaction of two
quasi-real photons. Here we focus on $e^+ e^-$ collisions, but we
briefly discuss also the case of production in proton-proton collisions
({\em hadroproduction}), via a gluon-initiated subprocess.

\section{Theoretical setup: photoproduction case}

The impact factor relevant for the process given in~(\ref{process})
reads~\cite{heavy}, at leading order \footnote{In Ref.~\cite{Celiberto:2017nyx}
  the factor $\sqrt{N_c^2-1}$ was forgotten; plots and tables in the present
  paper take it properly into account and, thus, overwrite the corresponding
  ones in Ref.~\cite{Celiberto:2017nyx}.}
\[
  d{\Phi}=\frac{\alpha\alpha_s e_Q^2}{\pi}\sqrt{N_c^2-1}
  \left[ m^2R^2+\vec P^2\left(z^2
    +{\overline z}^2\right)\right]d^2q\ dz\, ,
\]
\[
  R=\frac{1}{m^2+\vec q^{\, \,2}}-\frac{1}{m^2+(\vec q-\vec k)^2}\;,
  \;\;\;\;\;
  \vec P=\frac{\vec q}{m^2+\vec q^{\, \, 2}}+\frac{\vec k-\vec q}
       {m^2+(\vec q-\vec k)^2}\, .
\]
Here $\alpha$ and $\alpha_s$ denote the QED and QCD couplings, $N_c$ the
number of colors, $e_Q$ the electric charge of the heavy quark, $m$ its mass,
$z$ and ${\overline z}\equiv 1-z$ the longitudinal fractions
of the quark and antiquark produced in the same vertex and
$\vec k$, $\vec q$, $\vec k-\vec q$ the
transverse momenta with respect to the photons collision axis of Reggeized
gluon, produced quark and antiquark, respectively.

\begin{figure}[h]
\centering
\includegraphics[scale=0.34,clip]{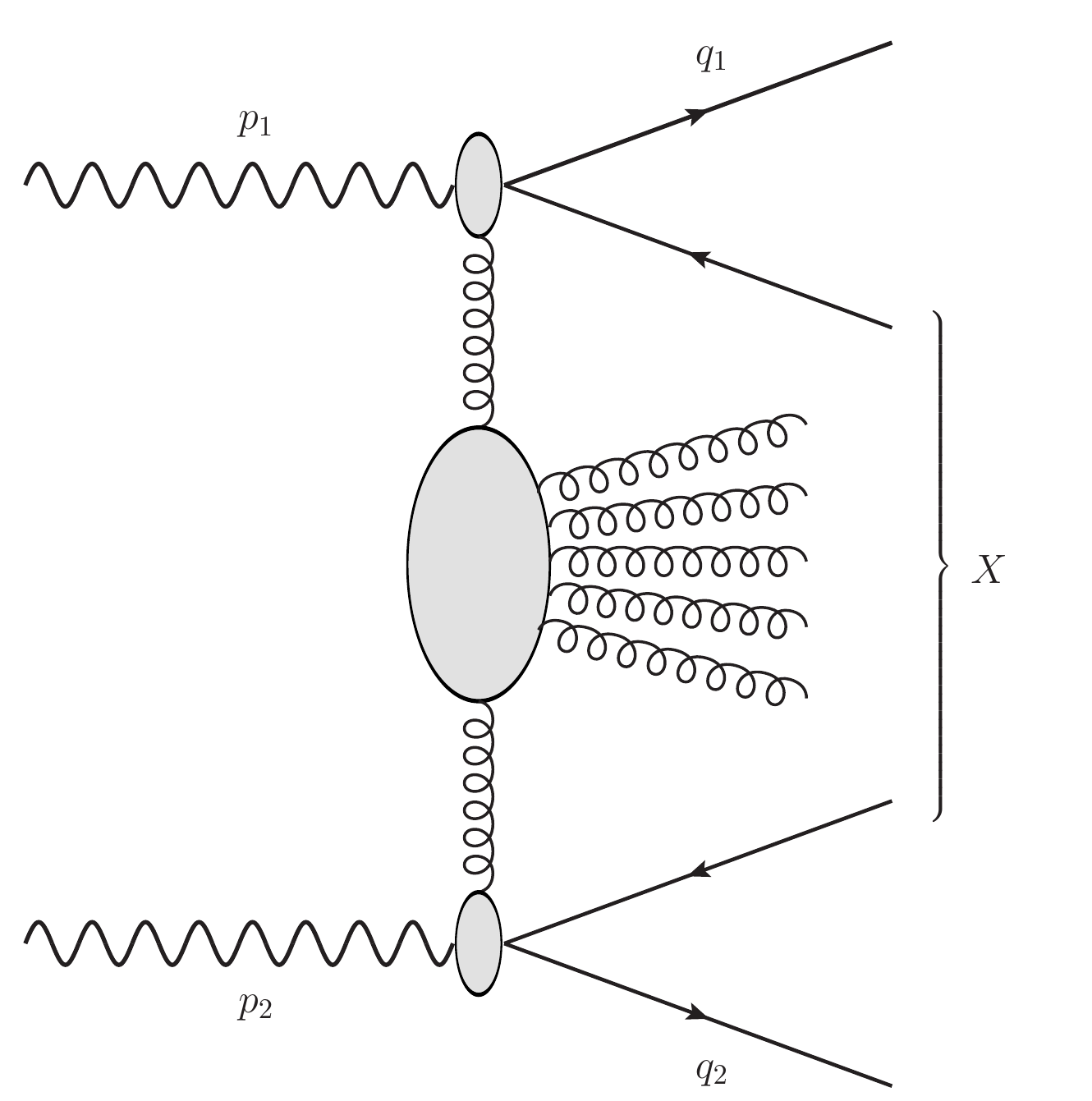}
\includegraphics[scale=0.51,clip]{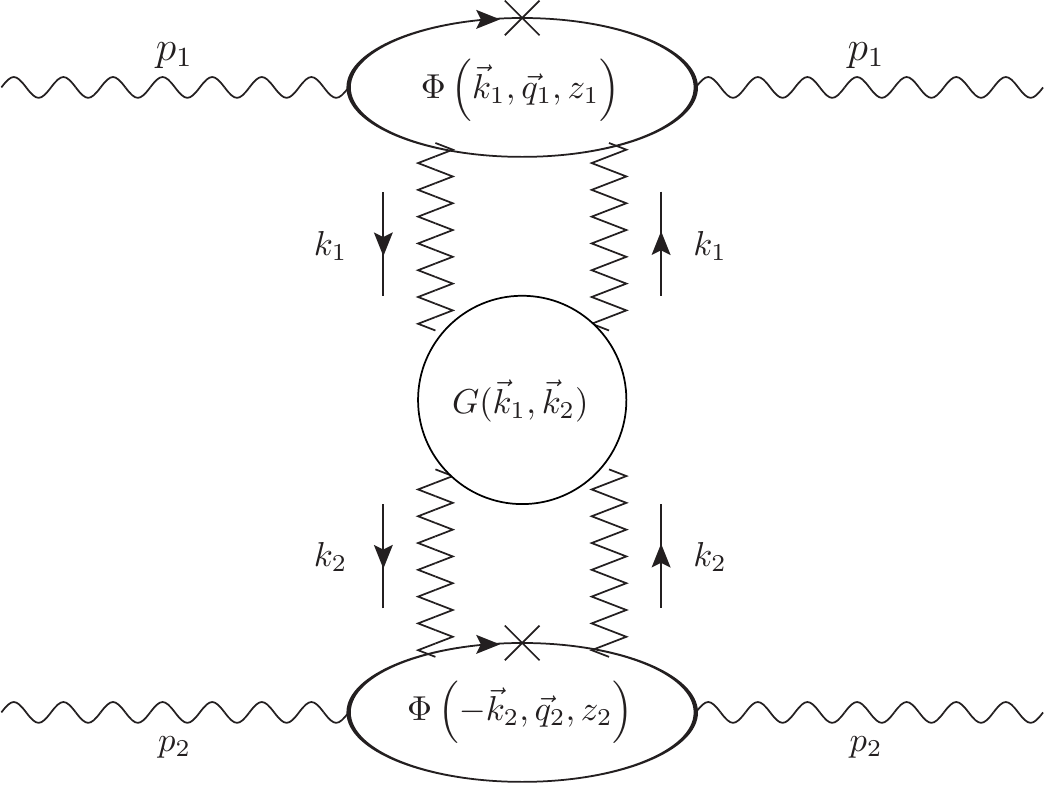}

\caption[]{(Left) Heavy-quark pair photoproduction. (Right) BFKL factorization:
  crosses denote the tagged quarks, whose momenta are not integrated over for
  getting the cross section.}
\label{fig:process}
\end{figure}
  
Similarly to the dihadron production processes (see~\cite{hh}), we have
\[
\frac{d\sigma_{\gamma\gamma}}{dy_1 dy_2 d|\vec q_1| d|\vec q_2| d\varphi_1
    d\varphi_2}
= \frac{1}{(2\pi)^2} \left[{\cal C}_0 +2\sum_{n=1}^\infty \cos(n\varphi) {\cal C}_n
\right]\;,
\]
where $\varphi = \varphi_1-\varphi_2-\pi$, with $\varphi_{1,2}$, $\vec q_{1,2}$
and $y_{1,2}$, respectively, the azimuthal angles, the transverse momenta
and the rapidities of the produced quarks. The ${\cal C}_n$ coefficients encode
the two leading-order impact factors and the NLA BFKL Green's function (see Fig.~\ref{fig:process}(right)). For brevity, the expression for ${\cal C}_n$ is not
presented here; it can be found in Ref.~\cite{Celiberto:2017nyx}, to which we
refer for all details concerning the present paper.

For the process initiated by $e^+e^-$ collisions, we must take into account the
flux of quasi-real photons $dn/dx$ emitted by each of the two colliding
particles. The cross section, differential in the rapidity gap
$\Delta Y$ between the two tagged heavy quarks, reads then
\[
\frac{d\sigma_{e^+e^-}}{d\left(\Delta Y\right)}=
\int_{q_{\rm min}}^{q_{\rm max}} d q_1\int_{q_{\rm min}}^{q_{\rm max}} d q_2
    \int_{-y_{\rm max}^{(1)}}^{y_{\rm max}^{(1)}} dy_1
    \int_{-y_{\rm max}^{(2)}}^{y_{\rm max}^{(2)}} dy_2\, \delta\left(y_1-y_2-\Delta Y\right)
\]
\beq{final}
\times \int_{e^{-\left(y_{\rm max}^{(1)}-y_1\right)}}^{1}\frac{dn_1}{dx_1}dx_1
\int_{e^{-\left(y_{\rm max}^{(2)}+y_2\right)}}^{1}\frac{dn_2}{dx_2}dx_2\,
d\sigma_{\gamma\gamma}\;,
\eeq
with $y_{\rm max}^{(1)}=\ln\sqrt{\frac{s}{m_1^2+\vec q^{\, \, 2}_1}}$ and
$y_{\rm max}^{(2)}=\ln\sqrt{\frac{s}{m_2^2+\vec q^{\, \, 2}_2}}$, where $s$ is
the squared center-of-mass energy of the colliding $e^+ e^-$ pair.
In the following we will present results for the \emph{integrated} azimuthal
coefficients $C_n$, defined through
\[
\frac{d\sigma_{e^+e^-}}{d\left(\Delta Y\right) d\varphi_1 d\varphi_2}
= \frac{1}{(2\pi)^2} \left[C_0 +2\sum_{n=1}^\infty \cos(n\varphi) C_n
\right]\;.
\]
 
\begin{table}[tb]
  \small
  \centering
  \caption{$C_0$ [pb] {\it vs.} $\Delta Y$ for $q_{\rm min} = 0$ GeV and
    $\sqrt{s} = 200$ GeV;
    $C$ stands for $\mu_R^2/(s_1 s_2)$, with $s_{1,2} = m_{1,2}^2 + \vec q_{1,2}^2$.}
\label{tab:C0-200GeV}
\begin{tabular}{r|lllllll}
\toprule
$\Delta Y$ & 
Box $q\bar{q}$ &
$\barr c \rm LLA \\ C = 1/2 \earr$ & 
$\barr c \rm LLA \\ C = 1 \earr$ &
$\barr c \rm LLA \\ C = 2 \earr$ &
$\barr c \rm NLA \\ C = 1/2 \earr$ &
$\barr c \rm NLA \\ C = 1 \earr$ & 
$\barr c \rm NLA \\ C = 2 \earr$ \\
\midrule
1.5 & 98.26  & 415.0(1.3) & 65.24(31) & 28.94(14) & 16.96(10) &11.237(73) &10.289(74)  \\
2.5 & 42.73  & 723.7(2.1) & 88.64(36) & 34.58(17) & 17.580(91)& 9.581(57) & 8.504(56)  \\
3.5 & 14.077 &1203.4(3.4) &113.33(43) & 39.01(16) & 18.522(92)& 7.989(43) & 6.637(36)  \\   
4.5 & 3.9497 &1851.6(5.0) &133.64(52) & 40.42(19) & 18.412(90)& 6.210(31) & 4.893(25)  \\
5.5 & 0.9862 &2559.4(7.1) &140.23(55) & 37.18(17) & 16.971(83)& 4.329(21) & 3.138(15)  \\
\bottomrule
\end{tabular}
\end{table}

\begin{table}[tb]
  \small
  \centering
  \caption{$C_0$ [pb] {\it vs.} $\Delta Y$ for $q_{\rm min} = 0$ GeV and
    $\sqrt{s} = 3$ TeV;
    $C$ stands for $\mu_R^2/(s_1 s_2)$, with $s_{1,2} = m_{1,2}^2 + \vec q_{1,2}^2$.}
\label{tab:C0-3TeV}
\setlength{\tabcolsep}{5.3pt}
\begin{tabular}{r|lllllll}
\toprule
$\Delta Y$ & 
Box $q\bar{q}$ &
$\barr c \rm LLA \\ C = 1/2 \earr$ & 
$\barr c \rm LLA \\ C = 1 \earr$ &
$\barr c \rm LLA \\ C = 2 \earr$ &
$\barr c \rm NLA \\ C = 1/2 \earr$ &
$\barr c \rm NLA \\ C = 1 \earr$ & 
$\barr c \rm NLA \\ C = 2 \earr$ \\
\midrule
1.5 & 280.98 &   10.893(49)$\cdot 10^3$ &  530.8(2.4)           &  195.54(88) &   99.57(89) & 58.34(58) & 52.17(58) \\
3.5 & 48.93  &   54.84(14) $\cdot 10^3$ & 1568.9(7.6)           &  439.4(2.1) &  184.5(1.1) & 65.22(50) & 54.38(47) \\
5.5 & 4.9819 &  254.88(57) $\cdot 10^3$ & 4409(19)              &  930.6(4.2) &  380.2(1.8) & 75.83(53) & 55.22(36) \\
7.5 & 0.4318 & 1041.7(2.1) $\cdot 10^3$ &10.921(44)$\cdot 10^3$ & 1743.1(8.3) &  756.3(3.5) & 81.94(44) & 51.48(27) \\
9.5 & 0.0323 & 3429.0(7.8) $\cdot 10^3$ &21.530(80)$\cdot 10^3$ & 2618(12)    & 1267.0(6.1) & 72.73(36) & 38.86(19) \\
10.5& 0.0081 & 5468(14)    $\cdot 10^3$ &26.23(10) $\cdot 10^3$ & 2761(12)    & 1443.3(7.2) & 59.97(30) & 29.21(14) \\
\bottomrule
\end{tabular}
\end{table}

\section{Numerical analysis}

We consider only the case of $c$-quark with mass $m=1.2$ GeV/$c^2$ and
fix $q_{\rm max}$ = 10 GeV and $q_{\rm min} = 0$, 1, 3 GeV.
We take $\sqrt{s} = 200$ GeV, as in LEP2, with $1 < \Delta Y < 6$, and 
$\sqrt{s} = 3$ TeV, as in the future $e^+e^-$ CLIC linear accelerator,
with $1 < \Delta Y < 11$.

In Tables~\ref{tab:C0-200GeV}-\ref{tab:C0-3TeV} we show pure LLA and NLA BFKL
predictions for $C_0$ with $q_{\rm min} = 0$ GeV and $\sqrt{s}$ = 200~GeV and
3~TeV, respectively, and compare them with the exclusive photoproduction of
a $c\bar c$ pair, given by two ``box'' diagrams. We see that at LEP2 energies
the ``box'' cross section dominates, but at CLIC energies BFKL takes over.

\begin{figure}[tb]
\centering

   \includegraphics[scale=0.32,clip]{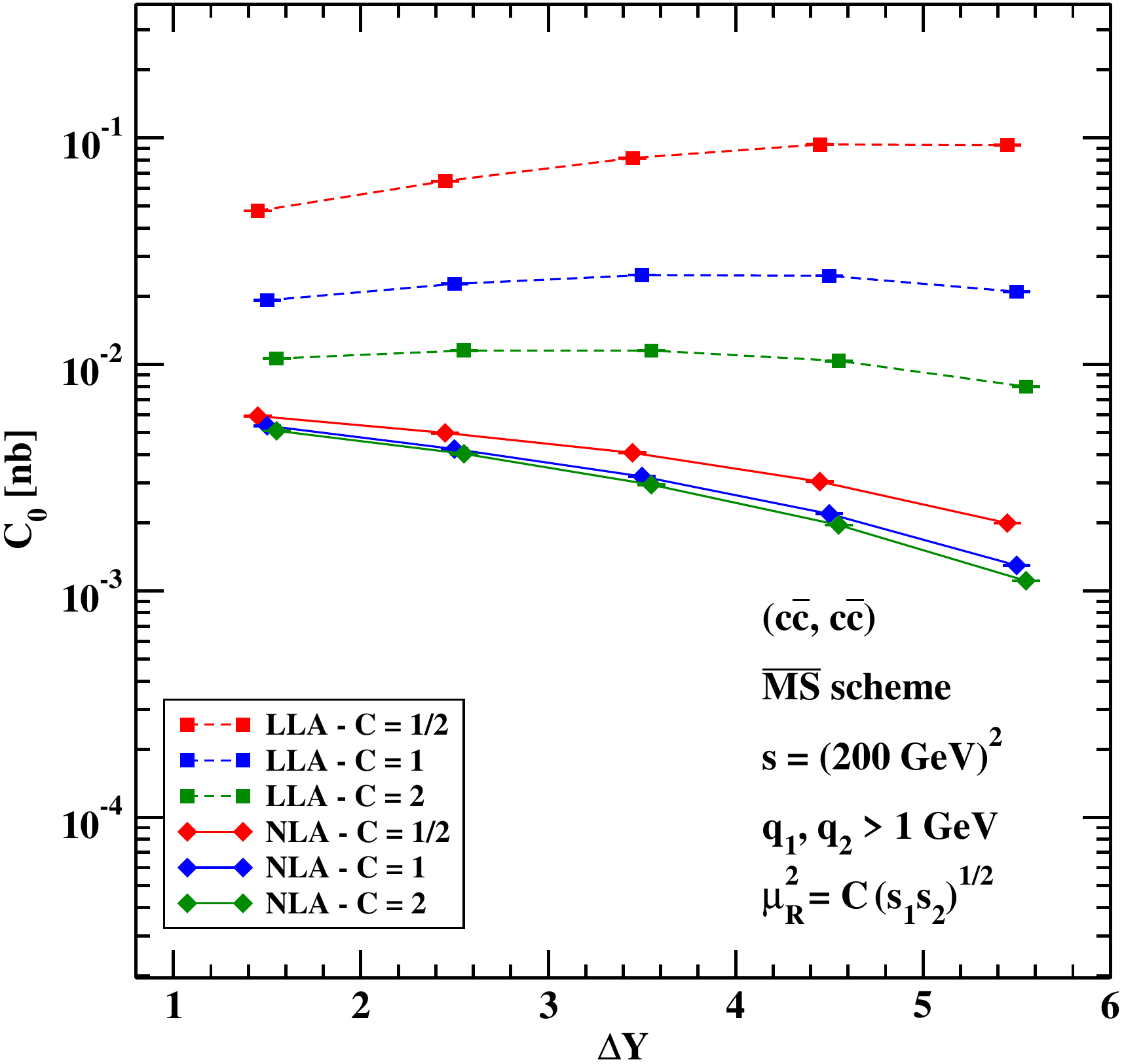}
   \includegraphics[scale=0.32,clip]{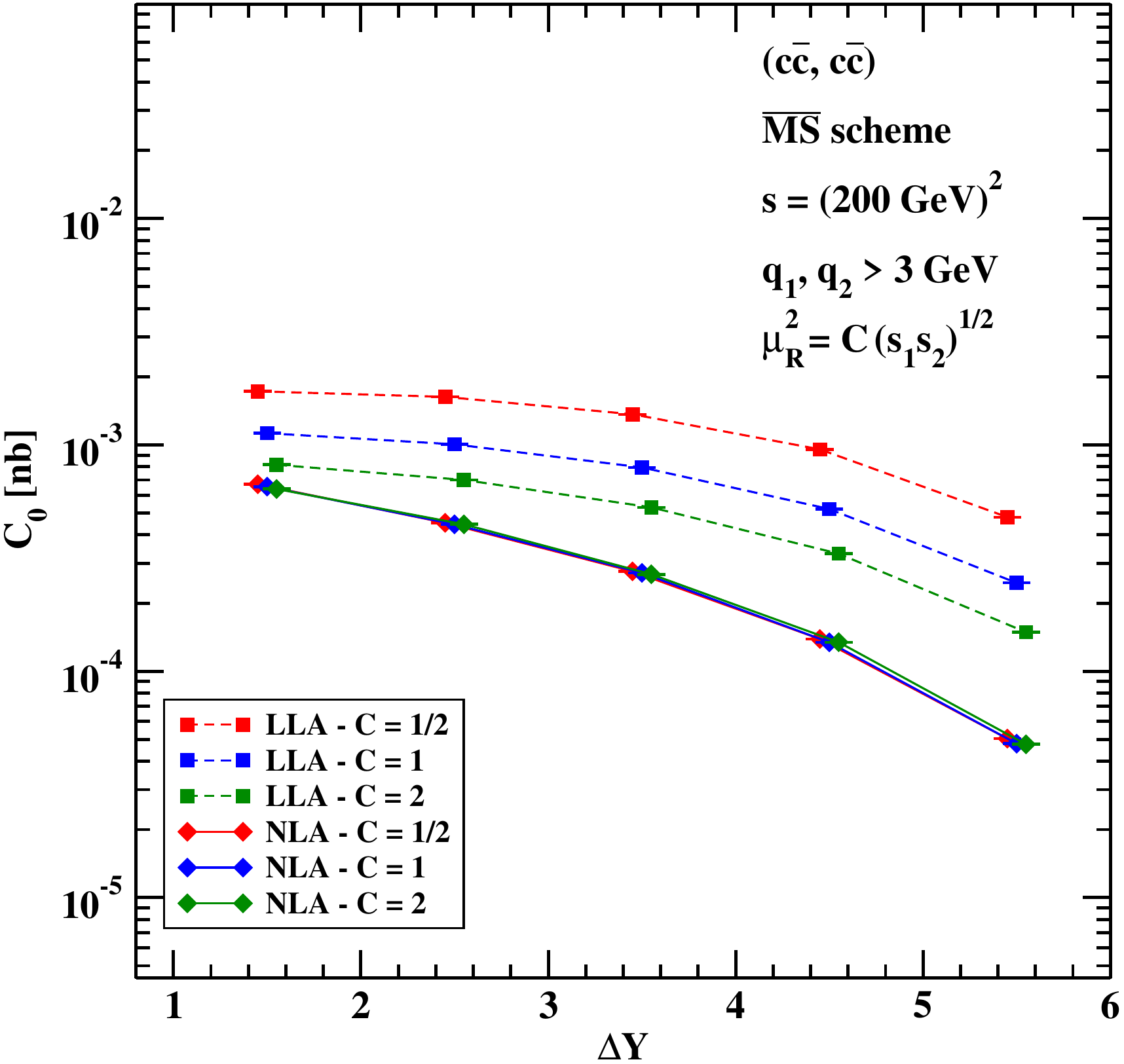}

   \includegraphics[scale=0.32,clip]{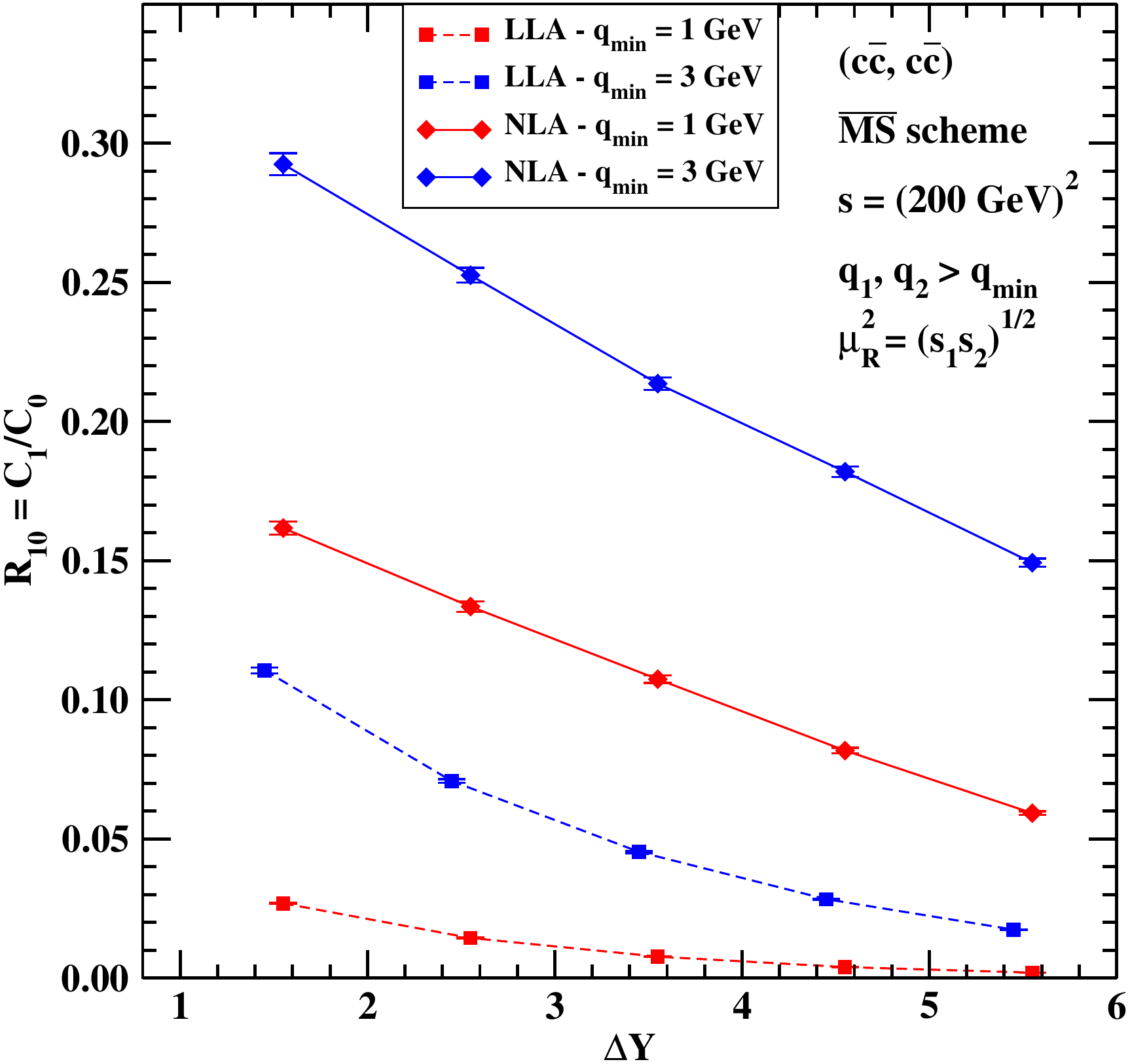}
   \includegraphics[scale=0.32,clip]{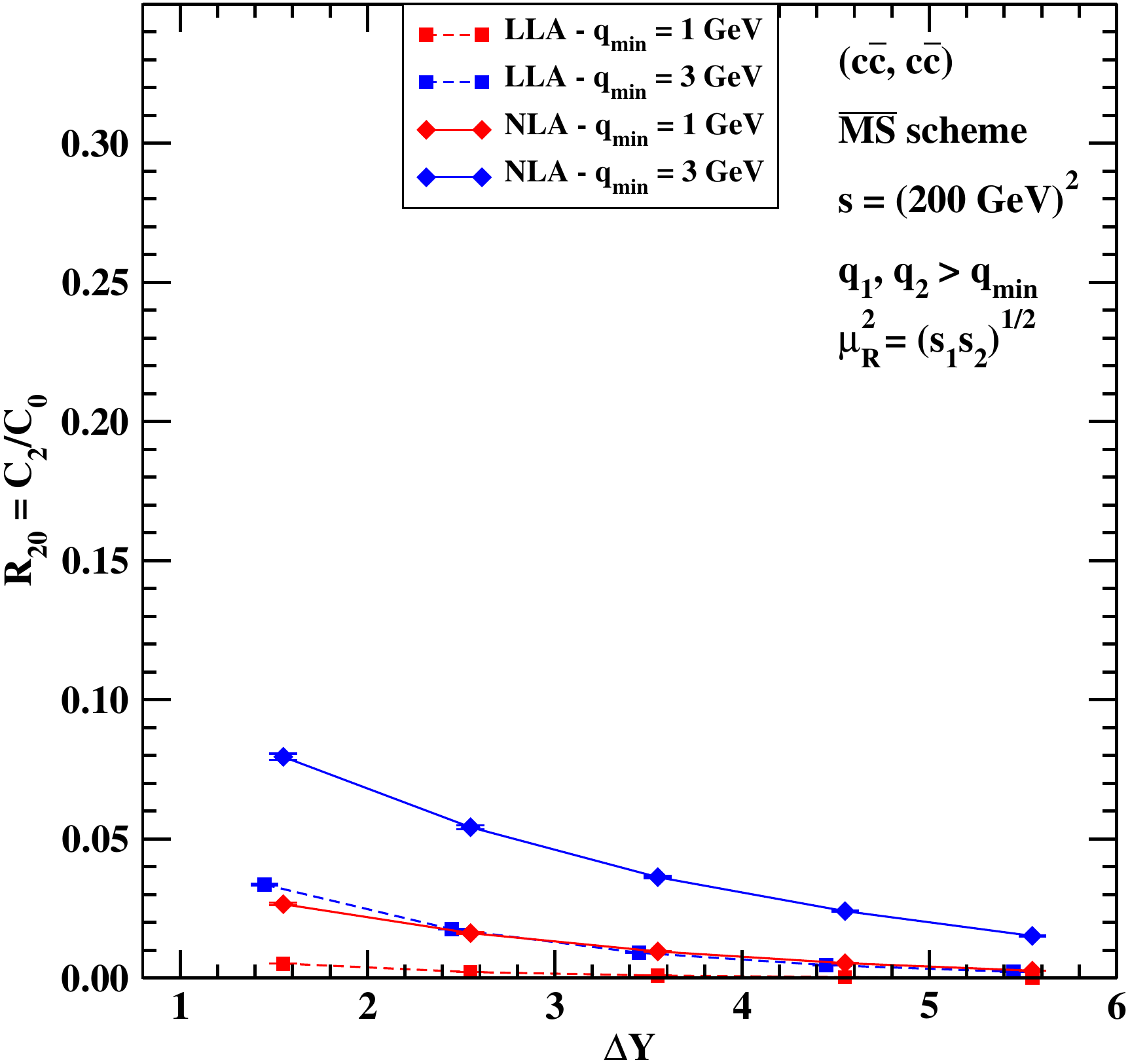}

   \includegraphics[scale=0.32,clip]{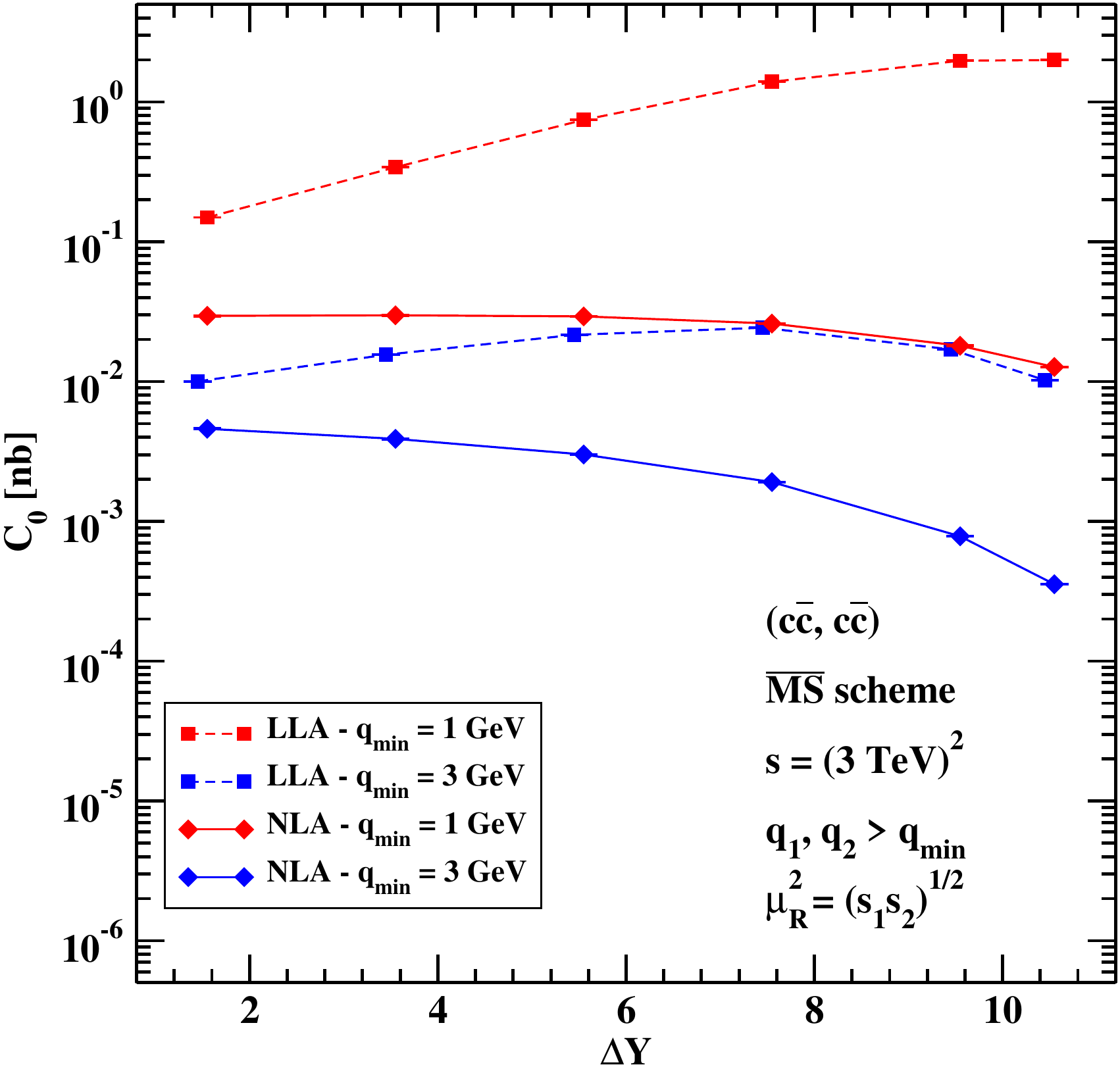}
   \includegraphics[scale=0.32,clip]{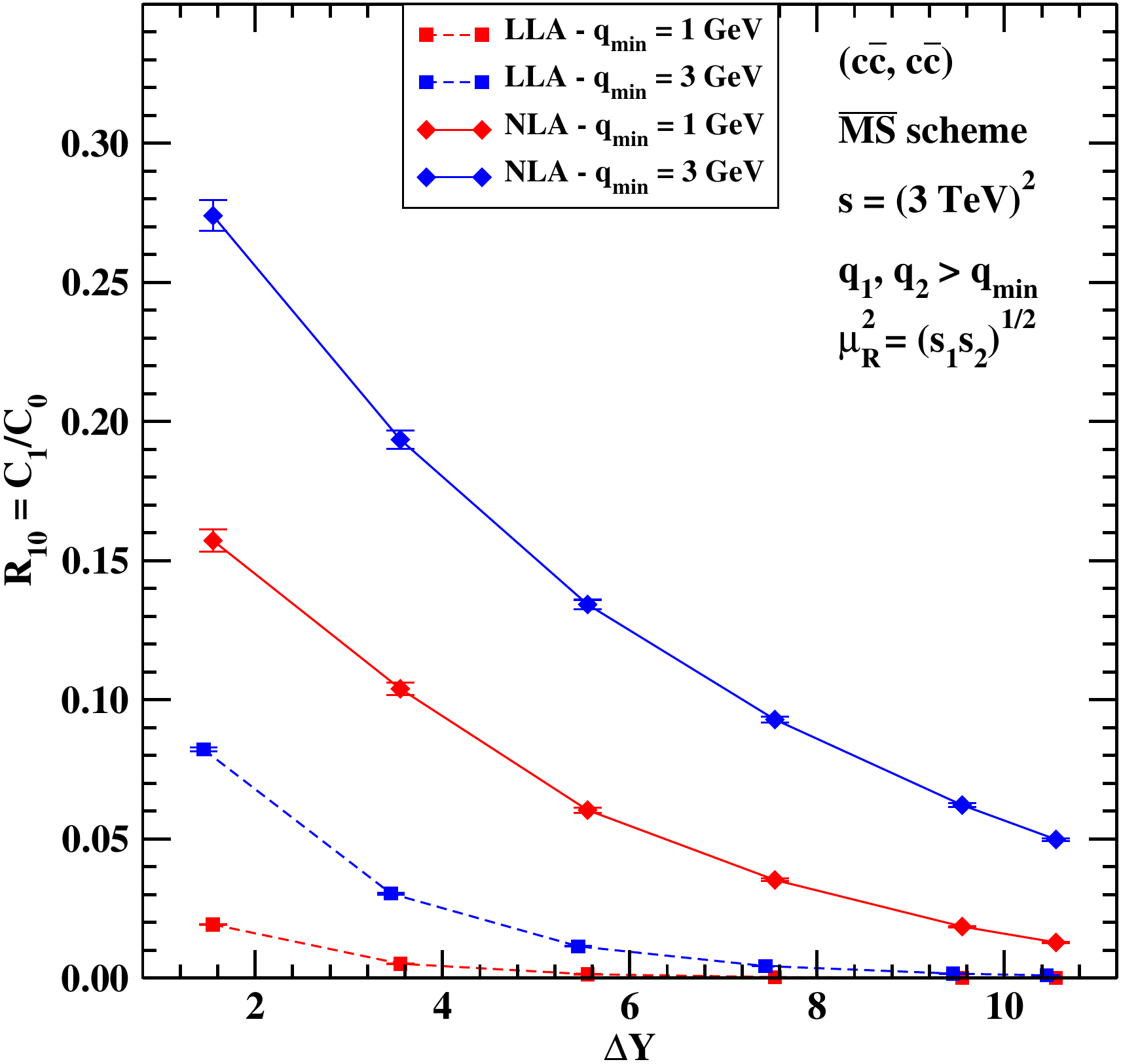}

   \caption{$\Delta Y$-dependence of $C_0$, $R_{10}$, and $R_{20}$ for
     $q_{\rm min} = 1$, 3 GeV, $\sqrt{s} = 200$ GeV and 3 TeV, and for different
     values of $C = \mu_R^2/\sqrt{s_1 s_2}$, with $s_{1,2} = m_{1,2}^2
     + \vec q_{1,2}^2$.}
   \label{fig}

\end{figure}

Results for $C_0$, $R_{10}$, and $R_{20}\equiv C_2/C_0$ with $q_{\rm min} = 1$,
3 GeV and $\sqrt{s} = 200$ GeV and 3 TeV are shown in Fig.~\ref{fig}.
We see that the cross section increases from LEP2 to CLIC energies and
decreases from LLA to NLA. Azimuthal correlations are in all cases much smaller
than one and decrease when $\Delta Y$ increases, as it must be due to the larger
emission of undetected partons. Moreover, the inclusion of NLA effects
increases the correlations, which can only be explained with the larger
suppression of $C_0$ with respect to $C_{1,2}$ when these effects are included.  

\section{Theoretical setup: hadroproduction case}

The hadroproduction case can be studied in a similar fashion as the
photoproduction one, with the role of photons played by gluons and
the photon flux replaced by the gluon parton distribution function in the
proton. The differential impact factor in this case takes the form
\[
\frac{d{\Phi}}{d^2q\ dz}=\frac{\alpha_s^2 \sqrt{N_c^2-1}}{2\pi N_c}\left[\left(m^2\left(R
  +\bar R\right)^2+\left(\vec  P+\vec {\bar P}\right)^2\left(z^2
+\bar z^2\right)\right)\right.
\left. -2\frac{N_c^2}{N_c^2-1}
\left(m^2 R \bar R+\vec  P\vec {\bar P}\left(z^2 +\bar z^2\right)
\right) \right]\;,
\]
\[
R=\frac{1}{m^2+\vec q^{\,2}}-\frac{1}{m^2+(\vec q-\vec k z )^2}\;,
  \;\;\;\;\;
\vec  P=\frac{\vec q}{m^2+\vec q^{\, 2}}+\frac{\vec k z-\vec q}
      {m^2+(\vec q-\vec k z)^2}\;,
\]
\[
\bar R=\frac{-1}{m^2+(\vec q-\vec k)^2}+\frac{1}{m^2+(\vec q-\vec k z )^2}\;,
\;\;\;\;\;
\vec{\bar P}=\frac{\vec k -\vec q}{m^2+(\vec k -\vec q)^2}
-\frac{\vec k z-\vec q}{m^2+(\vec q-\vec k z)^2}\;.
\]
Color and coupling prefactors enhance hadroproduction cross section
by some $10^3$ with respect to photoproduction, but photon flux $dn/dx$
dominates over $g(x)$ for $x\to 0$ and $x\to 1$ so that it is not easy to
estimate the size of the cross section without a detailed numerical analysis.

\end{document}